\documentclass[aps,pra,showpacs,twocolumn,balance,superscriptaddress]{revtex4-2}

\usepackage{subfigure}
\usepackage{amssymb}
\usepackage{amsmath}
\usepackage{graphicx}
\usepackage{epsfig}
\usepackage{graphicx}
\usepackage{epstopdf}
\usepackage{float}
\usepackage{hyperref} 
\usepackage{cleveref} 

\hypersetup{colorlinks = true, 
	linkcolor = blue, 
	urlcolor = blue,
	citecolor = blue}

\begin{document}
	
	\title{Single-photon scattering in a dissipative superconducting-qubit--SSH lattice hybrid}
	
	\author{X. X. Zhang}
	\affiliation{College of Physics and Materials Science, Tianjin Normal University, Tianjin
		300387, China}
	
	\author{J. Zhou}
	\email{zhouj966@tjnu.edu.cn}
	\affiliation{College of Physics and Materials Science, Tianjin Normal University, Tianjin
		300387, China}
	
  \author{X. Z. Zhang}
  \email{zhangxz@tjnu.edu.cn}
  \affiliation{College of Physics and Materials Science, Tianjin Normal University, Tianjin
	300387, China}
  \affiliation{Interdisciplinary center, Tianjin Normal University, Tianjin
	300387, China}

	\begin{abstract}
We study single-photon scattering in a Su--Schrieffer--Heeger (SSH) photonic lattice locally coupled to a superconducting qubit with tunable loss or gain. Working in the single-excitation sector, we derive an explicit real-space scattering formulation for the full energy-dependent scattering matrix $S(E)$ and identify how its eigenvalues encode coherent perfect absorption, amplification, and spectral singular behavior. The analytical results are benchmarked against time-domain wave-packet simulations, which reproduce the stationary scattering probabilities with high accuracy. We show that the SSH dimerization, the qubit-induced non-Hermitian self-energy, and the synthetic gauge phase cooperate to reshape the reflection and transmission spectra in a highly selective way. In particular, changing the dimerization can switch the system between transmission-dominated and reflection-dominated regimes, while the flux provides a direct handle on interference and symmetry-controlled response. We also find a robust loss--gain correspondence in the reflection landscape and show that the linewidth broadening is governed predominantly by the magnitude $|\gamma|$ of the non-Hermitian coupling. These results establish a compact and experimentally relevant framework for topological scattering in superconducting quantum networks.
\end{abstract}
	\maketitle
	
\section{Introduction}\label{sec:introduction}

Topological photonics and superconducting quantum circuits provide a natural setting for studying controllable light--matter interaction at the single-quantum level~\cite{zhou2008controllable,blaisCircuitQuantumElectrodynamics2021,youssefiTopologicalLatticesRealized2022,ozawaTopologicalPhotonics2019,hafeziImagingTopologicalEdge2013,kjaergaardSuperconductingQubitsCurrent2020,warnerCoherentControlSuperconducting2025,luTopologicalPhotonics2014a,belloUnconventionalQuantumOptics2019,kimQuantumElectrodynamicsTopological2021a,gaoQuantumTopologicalPhotonics2024,sauerTheoryIntrinsicPropagation2020,jalali2023topological}. The former supplies band topology, sublattice interference, and protected transport channels; the latter offers large tunability, strong coupling, and programmable dissipation. Bringing these ingredients together opens a route to scattering phenomena that are simultaneously topological, interferometric, and genuinely non-Hermitian.

For waveguide and lattice QED platforms, scattering is one of the most direct observables~\cite{yan2021quantum,dongWaveguideQEDDissipative2025,zhuSinglephotonScatteringGiantatom2025a,piasotskiDiagrammaticApproachScattering2021,poddubnyVanWaalsMaterials2025,RSinglephotonscattering2025}. In an SSH environment, the response of an incident photon already depends sensitively on dimerization, band geometry, and the phase structure of the Bloch spinor. Once a local superconducting qubit is added, the qubit generates an energy-dependent self-energy that can be reactive or dissipative, and can therefore reshape the scattering resonances rather than merely broaden them. This is especially relevant for superconducting implementations, where engineered loss, gain, and parametric control are not secondary corrections but standard experimental resources. A central question is then how topology, local non-Hermiticity, and flux-controlled interference cooperate in the same scattering problem~\cite{luDynamicalTopologyChiral2025,linMeasuringNonHermitianTopological2025,mondalSuSchriefferHeegerModelFundamentals2025,benderRealSpectraNonHermitian1998,riveroTimereversalinvariantScalingLight2019,fengNonreciprocalLightPropagation2011,ramezaniNonreciprocalLocalizationPhotons2018,khurginNonreciprocalPropagationNonreciprocal2020}.

Here we address that question for a single superconducting qubit locally coupled to an SSH lattice and threaded by a synthetic gauge phase. Our focus is not only on the reflection and transmission coefficients themselves, but on the full scattering matrix and its eigenchannelsl~\cite{geConservationRelationsAnisotropic2012,jinUnitaryScatteringProtected2022,mostafazadehInvisibilityPTSymmetry2013,chongPTsymmetryBreakingLaserabsorber2010,meierObservationTopologicalSoliton2016}. This viewpoint makes it possible to discuss coherent perfect absorption, amplification, and symmetry-controlled transport within a unified framework~\cite{mostafazadehSpectralSingularitiesComplex2009,geContrastingEigenvalueSingularvalue2017,longhiPTSymmetricLaser2010,wanTimeReversedLasingInterferometric2011}. Using a real-space matching approach in the single-excitation sector, we derive closed-form expressions for the stationary scattering amplitudes and then verify them by direct wave-packet dynamics.

Three physical conclusions emerge clearly. First, the SSH dimerization can switch the device between sharply different transport regimes, including nearly perfect transmission and nearly perfect reflection. Second, the synthetic flux reorganizes the interference between the two sublattice coupling paths and thereby reshapes the scattering landscape in a controllable way. Third, the non-Hermitian qubit term enters as a local self-energy whose magnitude largely controls linewidth broadening, while its sign distinguishes attenuation from amplification and participates in a useful loss--gain correspondence.

The rest of the paper is organized as follows. Section~\ref{sec:general} introduces the general scattering formulation for the SSH lattice and fixes the Bloch conventions used throughout. Section~\ref{sec:qubit} specializes this framework to a dissipative or amplifying superconducting qubit coupled to the lattice and derives explicit formulas for the scattering amplitudes. And compares the stationary theory with wave-packet dynamics and discusses the topology- and flux-dependent transport response for left and right incidence. Section~\ref{conclusion} summarizes the main results and their implications for topological circuit-QED scattering devices.

\section{General Scattering Formalism}
	\label{sec:general}

The central object of our analysis is the single-photon scattering matrix $S$, which relates asymptotic incoming and outgoing amplitudes in the SSH leads. Because the lattice carries a two-sublattice internal structure, the scattering problem is richer than that of a simple monatomic chain: the Bloch eigenmode already contains a nontrivial relative phase between the $A$ and $B$ components, and a local qubit can interfere with both coupling paths.

In the absence of the scatterer, the SSH chain supports two Bloch bands. A bulk eigenstate can be written as
\begin{equation}
	\psi _{k}(n)=
	\begin{pmatrix}
		u_{A,k}(n) \\ 
		u_{B,k}(n)
	\end{pmatrix}
	e^{ikn},
\end{equation}
where $n$ labels the unit cell and the spinor $(u_{A,k},u_{B,k})^{T}$ is fixed by the sublattice structure. Once a local impurity or qubit is attached, the incoming and outgoing amplitudes are related by
\begin{equation}
	\begin{pmatrix}
		b_{L} \\ 
		b_{R}
	\end{pmatrix}
	=S
	\begin{pmatrix}
		a_{L} \\ 
		a_{R}
	\end{pmatrix}
	,\qquad 
	S=
	\begin{pmatrix}
		r_{L} & t_{R} \\ 
		t_{L} & r_{R}
	\end{pmatrix},
\end{equation}
with $a_{L(R)}$ and $b_{L(R)}$ denoting the incoming and outgoing amplitudes from the left (right). When several propagating internal channels are retained, each block of $S$ may itself be matrix valued.

\subsection{Eigenvalues of the scattering matrix and extreme scattering channels}

For a non-Hermitian scattering problem, it is often more informative to examine the eigenchannels of $S$ than the raw matrix elements. If
\begin{equation}
	S|\psi _{\alpha }\rangle =\lambda _{\alpha }|\psi _{\alpha }\rangle ,
\end{equation}
then $\lambda_{\alpha}$ gives the complex response of the coherent input state $|\psi_{\alpha}\rangle$. The corresponding intensity ratio is
\begin{equation}
	\Theta _{\alpha }=\frac{\langle \psi _{\alpha }|S^{\dagger }S|\psi _{\alpha }\rangle }{\langle \psi _{\alpha }|\psi _{\alpha }\rangle }=|\lambda _{\alpha }|^{2},
\end{equation}
while $\arg \lambda_{\alpha}$ is the associated scattering phase.

Two limiting situations are especially important. When $\lambda_{\alpha}=0$, the outgoing wave in that eigenchannel vanishes and the system acts as a coherent perfect absorber. When $|\lambda_{\alpha}|$ diverges, the same framework signals a lasing or spectral-singularity threshold. Between these limits, $|\lambda_{\alpha}|<1$ corresponds to net attenuation and $|\lambda_{\alpha}|>1$ to net amplification.

In the present SSH--qubit problem, the eigenvalue structure of $S$ is shaped by three ingredients at once: the SSH Bloch geometry, the qubit-induced non-Hermitian self-energy, and the synthetic flux $\Phi$. The interplay of these ingredients determines where absorption, amplification, or strongly asymmetric interference can occur, and it provides a compact way to organize the scattering physics discussed below.

\subsection{Bloch solutions of the SSH chain}

We begin with the standard Su--Schrieffer--Heeger (SSH) tight-binding model
with two sublattices, $A$ and $B$, per unit cell. In real space, the
(Hermitian) SSH Hamiltonian reads 
\begin{equation}
	H_{\mathrm{SSH}}=\sum_{n}\left( t_{1}a_{n}^{\dag }b_{n}+t_{2}a_{n+1}^{\dag
	}b_{n}+\text{H.c.}\right) ,  \label{H_SSH_real}
\end{equation}%
where $a_{n}^{\dag }$ ($b_{n}^{\dag }$) creates a particle on sublattice $A$
($B$) of unit cell $n$; $t_{1}=-t(1+\delta )$ and $t_{2}=-t(1-\delta )$
represent the intracell and intercell hopping amplitudes, respectively, where%
$\delta $ is the dimerization parameter controlling the degree of
bond alternation.

By introducing the Fourier transforms 
\begin{equation}
	a_{n}=\frac{1}{\sqrt{N}}\sum_{k}a_{k}e^{ikn},\qquad b_{n}=\frac{1}{\sqrt{N}}%
	\sum_{k}b_{k}e^{ikn},
\end{equation}%
with the convention that the Brillouin zone is $k\in (-\pi ,\pi ]$, the
Hamiltonian in the momentum basis is 
\begin{equation}
	H_{\mathrm{SSH}}=\sum_{k}%
	\begin{pmatrix}
		a_{k}^{\dag } & b_{k}^{\dag }%
	\end{pmatrix}%
	\,H(k)\,%
	\begin{pmatrix}
		a_{k} \\ 
		b_{k}%
	\end{pmatrix}%
	,
\end{equation}%
where the Bloch Hamiltonian is a $2\times 2$ matrix 
\begin{equation}
	H(k)=%
	\begin{pmatrix}
		0 & h(k) \\ 
		h^{\ast }(k) & 0%
	\end{pmatrix}%
	,\qquad h(k)=t_{1}+t_{2}e^{-ik}.  \label{Hk}
\end{equation}%
The eigenvalue equation, $\det [H(k)-EI]=0$, gives $E^{2}=|h(k)|^{2}$, hence
the two bands are 
\begin{equation}
	E_{s}(k)=s\,|h(k)|,\qquad s=\pm ,  \label{eq:SSH_dispersion}
\end{equation}%
with 
\begin{equation}
	|h(k)|^{2}=t_{1}^{2}+t_{2}^{2}+2t_{1}t_{2}\cos k.  \label{eq:h_modsq}
\end{equation}%
The band gap at the Brillouin-zone point $k=\pi $ is $\Delta =2|t_{2}-t_{1}|$%
. Let $h(k)=|h(k)|e^{i\theta (k)}$ define the phase $\theta (k)=\arg h(k)$.
For $E\neq 0$, the eigenvector components satisfy $h(k)u_{B,k}=Eu_{A,k}$ and 
$h^{\ast }(k)u_{A,k}=Eu_{B,k}$. Thus, for the band $s=\pm $, one finds the
ratio 
\begin{equation}
	\frac{u_{B,k}^{(s)}(n)}{u_{A,k}^{(s)}(n)}=\frac{E_{s}(k)}{h(k)}=s\,\frac{%
		|h(k)|}{h(k)}=s\,e^{-i\theta (k)}.  \label{eq:ratio}
\end{equation}%
A convenient normalized choice of Bloch spinors is 
\begin{equation}
	u_{s}(k)=\frac{1}{\sqrt{2}}%
	\begin{pmatrix}
		1 \\[4pt] 
		s\,e^{-i\theta (k)}%
	\end{pmatrix}%
	,\qquad \langle u_{s}(k)\!,u_{s}(k)\rangle =1.  \label{eq:bloch_spinor}
\end{equation}%
The corresponding real-space Bloch wave (for unit cell $n$) is 
\begin{equation}
	\psi _{s,k}(n)=e^{ikn}u_{s}(k).  \label{eq:bloch_realspace}
\end{equation}%
Note that the chiral (sublattice) symmetry $\sigma _{z}H(k)\sigma _{z}=-H(k)$
maps $u_{s}(k)$ to $u_{-s}(k)$ by a sign flip on one sublattice, $%
u_{-s}(k)=\sigma _{z}u_{s}(k)$.

\subsection{Group velocity and flux normalization}\label{sec:group_velocity}

To connect the stationary scattering amplitudes with the time-domain wave-packet simulations, it is useful to normalize each propagating Bloch mode by its flux rather than by its probability density. For band index $s=\pm$, the group velocity is
\begin{equation}
	v_{s}(k)=\frac{dE_{s}(k)}{dk}=s\frac{d|h(k)|}{dk}=-s\frac{t_{1}t_{2}\sin k}{|h(k)|},  \label{eq:group_velocity}
\end{equation}
which fixes the propagation direction and the carried current of the mode.

For a plane wave of amplitude $A$, the associated flux is proportional to $|A|^{2}v_{s}(k)$. A unit-flux incoming state is therefore obtained by rescaling the Bloch spinor by $1/\sqrt{|v_{s}(k)|}$:
\begin{equation}
	\Phi_{s,k}(n)=\frac{1}{\sqrt{|v_{s}(k)|}}\,e^{ikn}u_{s}(k).
\end{equation}
With this convention, reflection and transmission probabilities extracted from the stationary problem can be compared directly with the long-time weights of an evolved wave packet. The same normalization also makes the role of the SSH band geometry transparent: once the central momentum is fixed, the packet velocity is set by $v_{s}(k)$, whereas all deviations from free propagation are encoded in the local scattering amplitudes.

\subsection{Single-excitation ansatz and general scattering solution}

In the single-excitation subspace, the joint state of the superconducting
qubit and the SSH photon lattice can be written as 
\begin{equation}
	\Psi _{k}=u_{e}|e\rangle |\mathrm{vac}\rangle
	+\sum_{x}[u_{A,k}(n)a_{n}^{\dag }+u_{B,k}(n)b_{n}^{\dag }]|g\rangle |\mathrm{%
		vac}\rangle ,  \label{eq:ansatz_single_exc}
\end{equation}%
where $|\mathrm{vac}\rangle $ denotes the vacuum of the SSH lattice, $u_{e}$
is the excitation amplitude of the qubit, and $u_{A,k}(n)$, $u_{B,k}(n)$ are
the single-photon amplitudes on sublattices $A$ and $B$ of unit cell $n$ at
a crystal momentum (central momentum) $k$.

The stationary scattering problem is set by the time-independent Schr\"{o}dinger equation 
\begin{equation}
	H\,\Psi _{k}=E_{s}(k)\,\Psi _{k},  \label{eq:schrodinger}
\end{equation}%
with $H$ as the total Hamiltonian (SSH leads + local qubit scatterer) and $%
E_{s}(k),s=\pm $ as the single-photon energy within the upper ($+$) or lower
($-$) band. Away from the scattering region, the amplitudes satisfy the bulk
difference equations, and their general Bloch solutions reproduce the
two-component spinor structure of the SSH bands.

Consequently, for a single-photon incoming from the left in band $s$ with
quasi-momentum $k$, the asymptotic form of the amplitudes can be written in
the most general form. For compactness, we first present the
propagating-channel contribution for scattering at energy $E_{s}(k)$. The A-
and B-sublattice amplitudes admit the asymptotic representation 
\begin{subequations}
	\begin{align}
		u_{A,k}(n)& =%
		\begin{cases}
			e^{ikn}+r_{L}^{(s)}e^{-ikn}, & n\ll 0, \\[4pt] 
			t_{L}^{(s)}e^{ikn}, & n\gg 0,%
		\end{cases}
		\label{eq:uA_asymp} \\[4pt]
		u_{B,k}(n)& =%
		\begin{cases}
			s\,e^{-i\theta (k)}e^{ikn}+s\,r_{L}^{(s)}e^{i\theta (k)}e^{-ikn}, & n\ll 0,
			\\[4pt] 
			s\,t_{L}^{(s)}e^{-i\theta (k)}e^{ikn}, & n\gg 0,%
		\end{cases}
		\label{eq:uB_asymp}
	\end{align}%
	where $r_{L}^{(s)}$ and $t_{L}^{(s)}$ are the reflection and transmission
	amplitudes for an incident Bloch wave in band $s$ coming from the left.
	
	To determine the amplitudes $r_{L}^{(s)}$, $t_{L}^{(s)}$, and the qubit
	amplitude $u_{e}$, we impose the discrete Schr\"{o}dinger equation %
	\eqref{eq:schrodinger} at every lattice site inside the finite scattering
	region, including the site(s) to which the qubit is coupled, and the qubit
	eigenequation itself. Denoting collectively by $\{y\}$ the small set of
	sites that form the scattering region, these local equations provide a
	finite linear system of the form 
\end{subequations}
\begin{equation}
	\mathcal{M}(E)\,\mathbf{x}=\mathbf{b},  \label{eq:matching_linear_system}
\end{equation}%
where the unknown vector $\mathbf{x}=(u_{e},\{u_{A,k}(y)\},\{u_{B,k}(y)\},\{%
\alpha _{\mathrm{ev}}\})^{T}$ collects the internal amplitudes of the
scatterer and any amplitudes of evanescent channels $\{\alpha _{\mathrm{ev}%
}\}$ excited near the scatterer; $\mathbf{b}$ encodes the incoming wave (the
known incident amplitudes at the left asymptote); and $\mathcal{M}(E)$ is an
energy-dependent matrix determined by the SSH hoppings, the qubit energy,
and its coupling to the lead sites. Solving Eq.~%
\eqref{eq:matching_linear_system} yields the reflection and transmission
amplitudes, i.e., the entries of the scattering matrix $S(E)$ corresponding
to the chosen incoming channel.

The ansatz given by Eqs.~\eqref{eq:uA_asymp} and \eqref{eq:uB_asymp} is
completely general for propagating channels: it accounts for the two
sublattices (A/B) and the two bands ($s=\pm $) through the internal phase $%
\theta (k)$. Inter-band scattering and conversion between sublattice
components are naturally captured by nonzero off-diagonal elements of the
resulting $2\times 2$ transmission/reflection blocks.

When the scattering region couples strongly to the leads, evanescent modes
(complex-$k$ solutions at the same real energy $E$) must be retained in the
ansatz. These evanescent components appear explicitly in the finite set of
unknowns $\mathbf{x}$ in Eq.~\eqref{eq:matching_linear_system} and guarantee
that the local finite-difference equations are satisfied at the coupling
sites. From the solved amplitudes, one constructs the energy-resolved scattering
matrix in the usual block form 
\begin{equation}
	S(E)=%
	\begin{pmatrix}
		r_{L}(E) & t_{R}(E) \\ 
		t_{L}(E) & r_{R}(E)%
	\end{pmatrix}%
	,
\end{equation}%
where each block acts on the sublattice (A , B) internal space and may mix
band indices if multiple propagating channels are open at the same energy.

The same matching procedure applies when the qubit introduces dissipation or
an applied flux: dissipation enters via a complex self-energy in the qubit
eigenequation (a non-Hermitian term), and the flux modifies the hopping
phases and hence the Bloch phase $\theta (k)$. Both of these appear in $%
\mathcal{M}(E)$ and therefore alter $r_{L}^{(s)}$ and $t_{L}^{(s)}$.

In the following sections, we will employ this general scattering ansatz
both to derive analytic expressions in limiting regimes (weak coupling,
single open channel, resonant approximation) and to validate these
expressions against time-dependent wave-packet simulations where Gaussian
packets centered at momentum $k$ are launched and the reflected/transmitted
probabilities are extracted and compared to $|r_{L}^{(s)}|^{2}$ and $%
|t_{L}^{(s)}|^{2}$ obtained from the matching procedure.

\section{Scattering from a dissipative (or amplifying) superconducting qubit
	coupled to the SSH chain}
	\label{sec:qubit}

\begin{figure}[t]
	\centering
	\includegraphics[bb=32 101 917 466, width=13 cm,clip,width=1\linewidth]{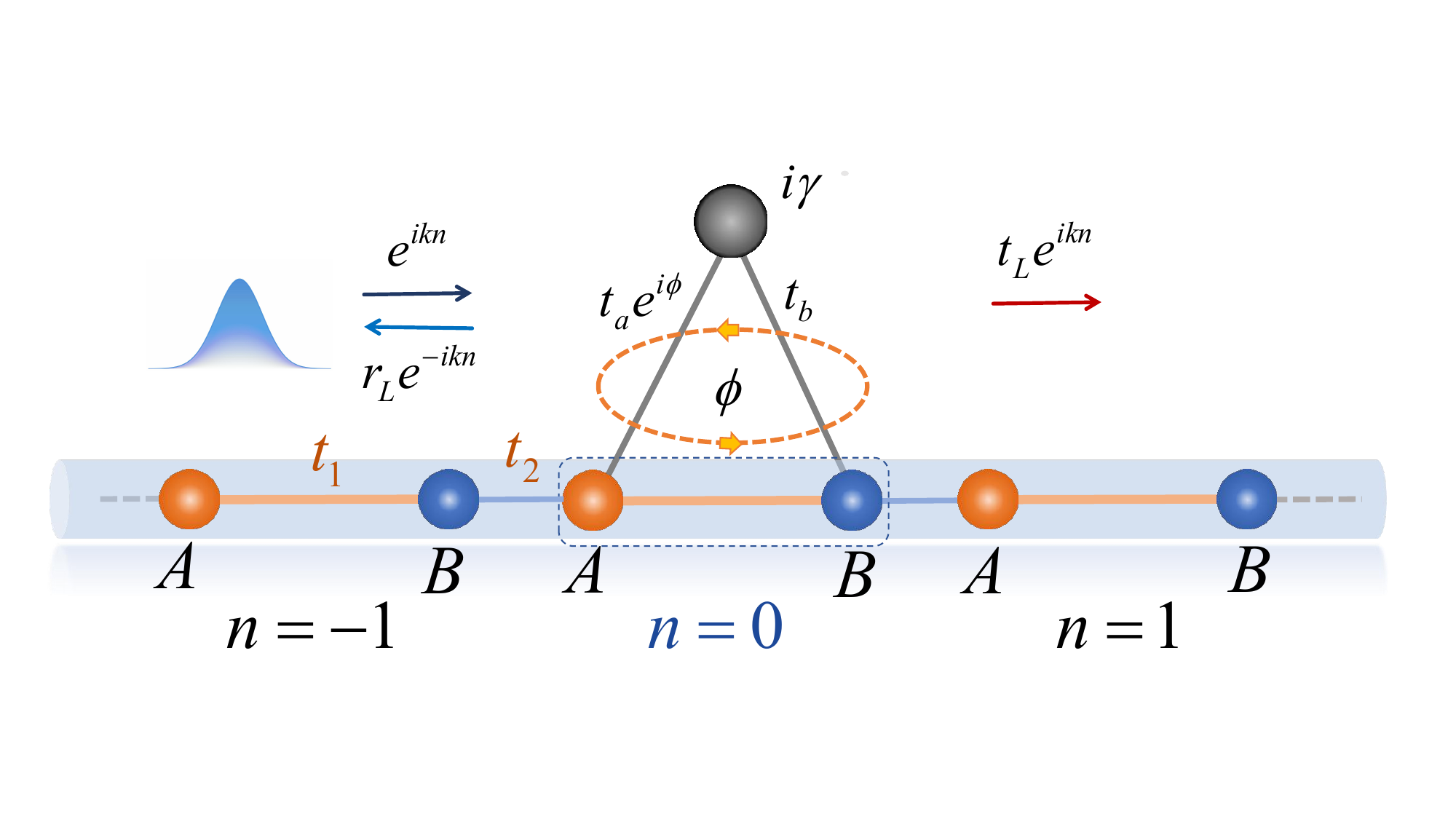}
	\caption{Schematic illustration of a dissipative (or amplifying) superconducting qubit embedded in the center of the SSH lattice. Superconducting qubit simultaneously coupled to both sublattices $A$ and $B$ of the unit
		cell at position $n=0$.}
	\label{fig:1qubit-AB1}
\end{figure}

We now consider a dissipative (or amplifying) superconducting qubit embedded
in the center of the SSH lattice, acting as a scattering impurity. The qubit is coupled locally to both sublattices $A$ and $B$ of the unit
cell at position $n=0$, shown in Fig.~\ref{fig:1qubit-AB1} . The ground and excited states of the qubit are
denoted by $|g\rangle $ and $|e\rangle $, respectively.

\subsection{Hamiltonian and physical realization}

The qubit Hamiltonian reads 
\begin{equation}
	H_{q}=i\gamma |e\rangle \langle e|,  \label{eq:Hq}
\end{equation}%
where the non-Hermitian term $i\gamma $ describes gain ($\gamma >0$) or dissipation ($\gamma <0$). Physically, such an effective model can be
realized by coupling a superconducting transmon or flux qubit to a
controlled decay channel or parametric amplifier, thereby introducing
tunable loss or gain into the local excitation.

The interaction between the qubit and the SSH chain is modeled as 
\begin{equation}
	H_{\mathrm{int}}= \left( t_{a}e^{i\Phi} a_{0}^{\dagger} + t_{b}
	b_{0}^{\dagger} \right)\sigma^{-} + \mathrm{H.c.},  \label{eq:Hint}
\end{equation}
where $t_{a}$ ($t_{b}$) denotes the coupling strength to the $A$ ($B$)
sublattice, $\Phi$ is the artificial gauge flux, and $\sigma^{-}=|g\rangle%
\langle e|$ ($\sigma^{+}=|e\rangle\langle g|$) are the lowering and raising
operators of the qubit.

The total Hamiltonian of the coupled system is therefore 
\begin{align}
	H_{\mathrm{tot}}=& \;H_{\mathrm{SSH}}+H_{q}+H_{\mathrm{int}}  \notag \\
	=& \;\sum_{n}[t_{1}a_{n}^{\dagger }b_{n}+t_{2}a_{n+1}^{\dagger }b_{n}+%
	\mathrm{H.c.}]+i\gamma |e\rangle \langle e|  \notag \\
	& \;+[(t_{a}e^{i\Phi }a_{0}^{\dagger }+t_{b}b_{0}^{\dagger })\sigma ^{-}+%
	\mathrm{H.c.}].  \label{eq:Htot}
\end{align}

\subsection{Explicit construction of the linear system $M(E)\mathbf{x}=%
	\mathbf{b}$}

For a single-photon incoming from the left in band $s$ with quasi-momentum $%
k$, in the single-excitation subspace, the stationary Schr\"{o}dinger
equation $H_{\mathrm{tot}}|\Psi _{k}\rangle =E_{s}(k)|\Psi _{k}\rangle $
yields the coupled equations at $n=0$: 
\begin{align}
	(-E_{s}+i\gamma )u_{e}+t_{a}e^{-i\Phi }u_{A,k}(0)+t_{b}u_{B,k}(0)& =0,
	\label{eq:site_e} \\[4pt]
	E_{s}u_{A,k}(0)-t_{2}u_{B,k}(-1)-t_{1}u_{B,k}(0)-t_{a}e^{i\Phi }u_{e}& =0,
	\label{eq:site_A} \\[4pt]
	E_{s}u_{B,k}(0)-t_{1}u_{A,k}(0)-t_{2}u_{A,k}(1)-t_{b}u_{e}& =0.
	\label{eq:site_B}
\end{align}

Using the asymptotic boundary conditions defined in Sec.~II, we eliminate
the qubit amplitude $u_{e}$ and reduce the scattering problem to a $2\times
2 $ linear system for the unknown reflection and transmission amplitudes $%
\mathbf{x}=(r_{L}^{(s)},t_{L}^{(s)})^{T}.$

First, from the qubit equation (\ref{eq:site_e}) we obtain 
\begin{equation}
	u_{e}=-\frac{t_{a}e^{-i\Phi }u_{A,k}(0)+t_{b}u_{B,k}(0)}{-E_{s}+i\gamma }.
	\label{eq:ue_elim}
\end{equation}

For a single photon incident from the left lead, we substitute the
asymptotic Bloch-wave forms into the discrete Schr\"{o}dinger equations at
the coupling site. Recalling that $\theta =\theta (k)\equiv \arg h(k)$\
specifies the relative phase between the sublattice amplitudes of the SSH
Bloch eigenstate, the wavefunction amplitudes at the contact can thus be
written as 
\begin{align}
	u_{A,k}(0)& =1+r_{L}^{(s)},  \label{eq:uA0} \\
	u_{B,k}(0)& =st_{L}^{(s)}e^{-i\theta (k)},  \label{eq:uB0} \\
	u_{A,k}(1)& =t_{L}^{(s)}e^{ik},  \label{eq:uA1_uBm1} \\
	u_{B,k}(-1)& =se^{-i\theta (k)}e^{-ik}+sr_{L}^{(s)}e^{i\theta (k)}e^{ik}.
	\label{eq:uA_uBm_2}
\end{align}%
Insert (\ref{eq:ue_elim}) and (\ref{eq:uA0})--(\ref{eq:uA_uBm_2}) into the
lattice equations at $n=0$, Eqs.~(\ref{eq:site_A}) and (\ref{eq:site_B}).
After straightforward algebra the two linear equations for $r_{L}^{(s)}$ and 
$t_{L}^{(s)}$ can be written in matrix form 
\begin{equation}
	M(E_{s})%
	\begin{pmatrix}
		r_{L}^{(s)} \\ 
		t_{L}^{(s)}%
	\end{pmatrix}%
	=\mathbf{b}(E_{s}),  \label{eq:Meqb_x}
\end{equation}%
with 
\begin{equation}
	M(E_{s})=%
	\begin{pmatrix}
		a & b \\[6pt] 
		c & d%
	\end{pmatrix}%
	,\qquad \mathbf{b}(E_{s})=%
	\begin{pmatrix}
		m_{1} \\ 
		m_{2}%
	\end{pmatrix}%
	,
\end{equation}%
and the matrix elements given explicitly by 
\begin{align}
	a& =E_{s}-st_{2}e^{i(\theta (k)+k)}+\frac{t_{a}^{2}}{-E_{s}+i\gamma },
	\label{eq:M_a} \\[4pt]
	b& =-st_{1}e^{-i\theta (k)}+\frac{st_{a}t_{b}}{-E_{s}+i\gamma }e^{i(\Phi
		-\theta (k))},  \label{eq:M_b} \\[4pt]
	c& =-t_{1}+\frac{t_{a}t_{b}}{-E_{s}+i\gamma }e^{-i\Phi },  \label{eq:M_c} \\%
	[4pt]
	d& =sE_{s}e^{-i\theta (k)}-t_{2}e^{ik}+\frac{st_{b}^{2}}{-E_{s}+i\gamma }%
	e^{-i\theta (k)}.  \label{eq:M_d}
\end{align}%
The right-hand side vector is 
\begin{equation}
	m_{1}=-a^{*},\qquad m_{2}=-c  \label{eq:b1}
\end{equation}

Here the signs $\pm$ consistently correspond to the upper/lower band ($s=\pm$%
), and $\theta=\theta(k)$ is the Bloch phase defined by $h(k)=|h(k)|e^{i%
	\theta}$. The terms proportional to $1/(E_{\pm}+i\gamma)$ arise from virtual
excitation of the qubit and encode both reactive (energy-dependent) and
dissipative (through $\gamma$) effects; these are the self-energy
corrections produced by integrating out the qubit degree of freedom.


\textit{Solution and physical limits.} Provided $\det M\neq 0$, the solution is obtained by direct matrix inversion
or, equivalently, by Cramer's rule: 
\begin{equation}
	\begin{split}
		r_{L}^{(s)} &= \frac{m_{1}d - m_{2}b}{\Delta }, \quad 
		t_{L}^{(s)} = \frac{am_{2} - cm_{1}}{\Delta }, \\
		\Delta &\equiv ad - bc.
	\end{split}
	\label{eq:rt_cramer}
\end{equation}%
The reflection and transmission probabilities follow as 
\begin{equation}
	R_{L}^{(s)}=|r_{L}^{(s)}|^{2},\qquad T_{L}^{(s)}=|t_{L}^{(s)}|^{2}.
\end{equation}

Several remarks are in order: 1. Poles of the scattering amplitudes
correspond to $\Delta =0$. In the complex-energy plane such poles indicate
resonances (finite-lifetime quasi-bound states) whose positions are shifted
and broadened by $\gamma $. Real-axis poles (or divergences in the physical
sheet) signal the threshold of self-sustained emission (lasing) in the
presence of gain ($\gamma >0$).

2. Zeros of the numerator $m_{1}d-m_{2}b$ (for $r_{L}^{(s)}$) or $%
am_{2}-cm_{1}$ (for $t_{L}^{(s)}$) can produce perfect
reflection/transmission suppression. In particular, if an eigenchannel of $S$
has eigenvalue $\lambda =0$ the corresponding coherent input is fully
absorbed (CPA). In our parametrization CPA points appear when the linear
system admits a nontrivial input with vanishing output, equivalently when
the homogeneous problem for outgoing amplitudes has a solution.

3. The non-Hermitian parameter $\gamma $ breaks unitarity of $S$; thus in
general $R_{L}^{(s)}+T_{L}^{(s)}\neq 1$. Positive $\gamma $ can amplify it
(gain), negative $\gamma $ reduces total flux (loss).

4. The gauge phase $\Phi $ enters only through the interference factors $%
e^{\pm i\Phi }$ in the qubit-mediated self-energy terms; by tuning $\Phi $
one can control interference between A- and B-couplings and thereby break
reciprocity between left/right scattering.

The explicit formulas (\ref{eq:M_a})--(\ref{eq:b1}) are algebraic and can be
simplified for particular parameter limits (weak coupling $t_{a,b}\ll
t_{1,2} $, on-resonance $E_{s}\approx 0$, single-band regime, etc.).
Numerical evaluation of $r_{L}^{(s)}(E,\Phi )$ and $t_{L}^{(s)}(E,\Phi )$
based on these expressions is straightforward and will be compared with the time-domain results discussed below.

\subsection{Numerical illustrations and symmetry-controlled reflection landscapes}

\begin{figure*}[tbp]
	\centering
	\includegraphics[bb=28 27 995 432, width=18 cm, clip]{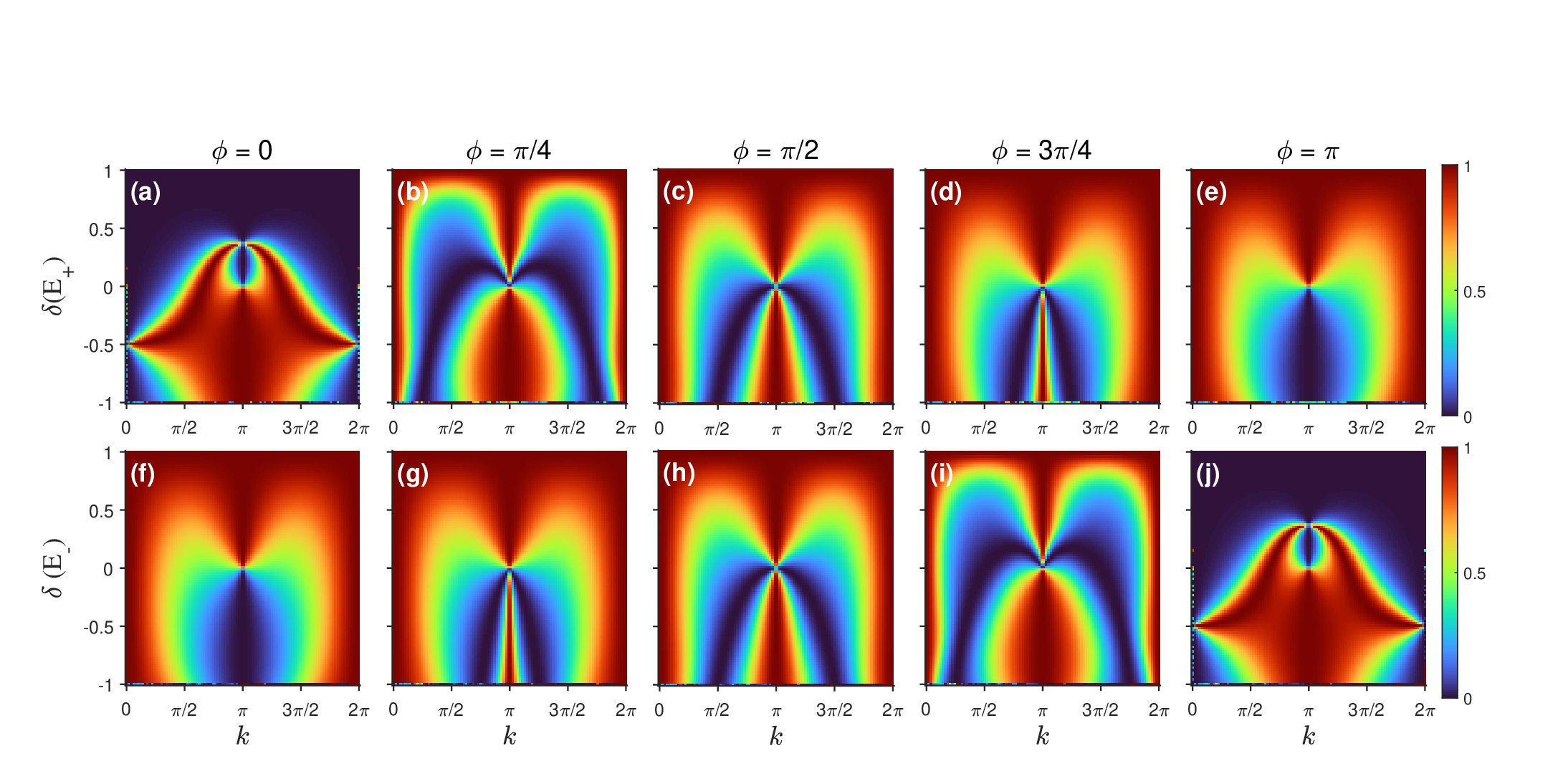}
	\caption{Reflection landscape for the AB geometry in the Hermitian limit, $t=1$ and $\gamma=0$.}
	\label{fig:R_vs_k_Phia}
\end{figure*}

\begin{figure*}[tbp]
	\centering
	\includegraphics[bb=28 27 995 432, width=18 cm, clip]{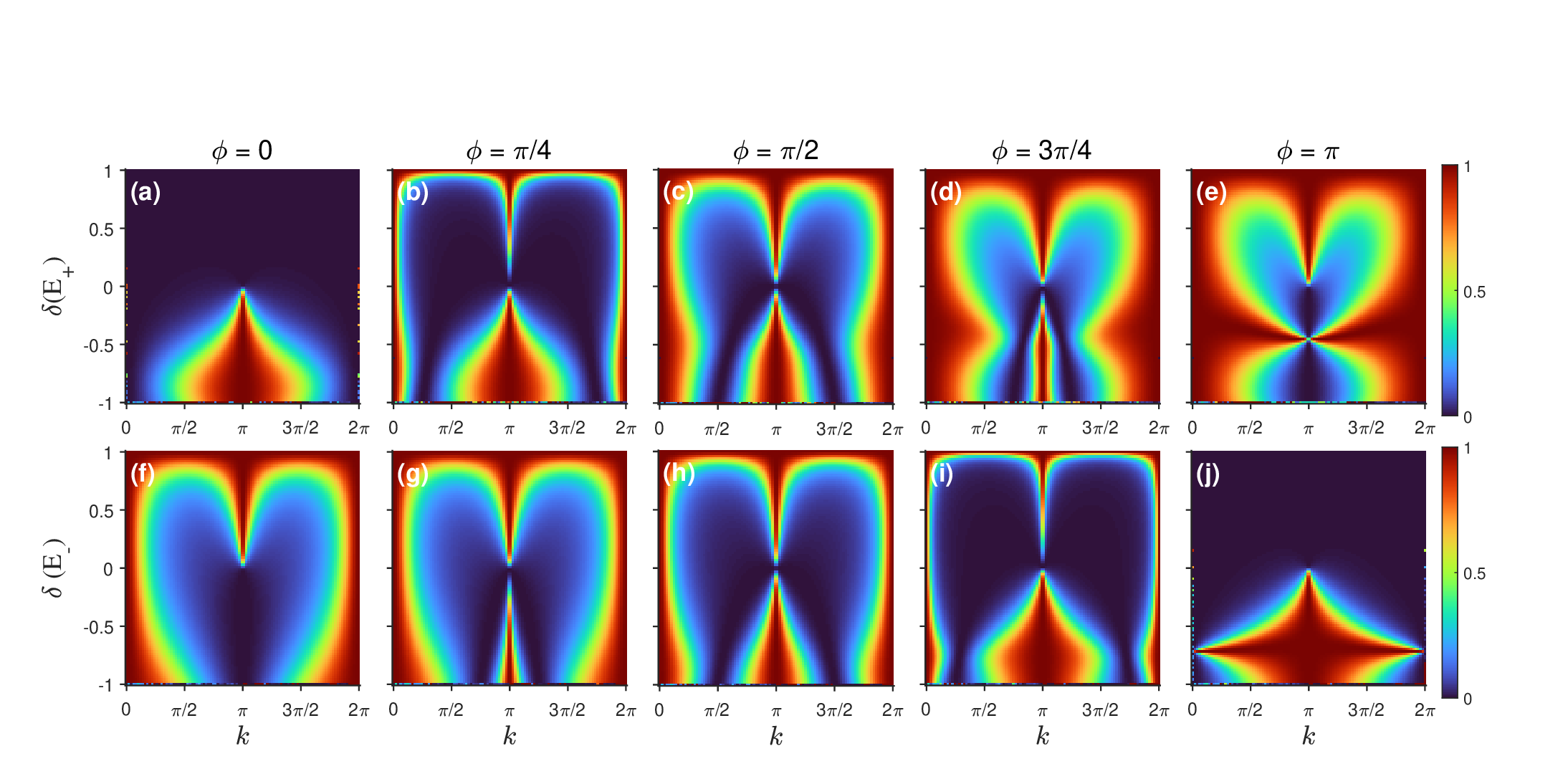}
	\caption{Reflection landscape for weak gain, $t=1$ and $\gamma=0.1$.}
	\label{fig:R_vs_k_Phib}
\end{figure*}

\begin{figure*}[tbp]
	\centering
	\includegraphics[bb=24 267 996 432, width=18 cm, clip]{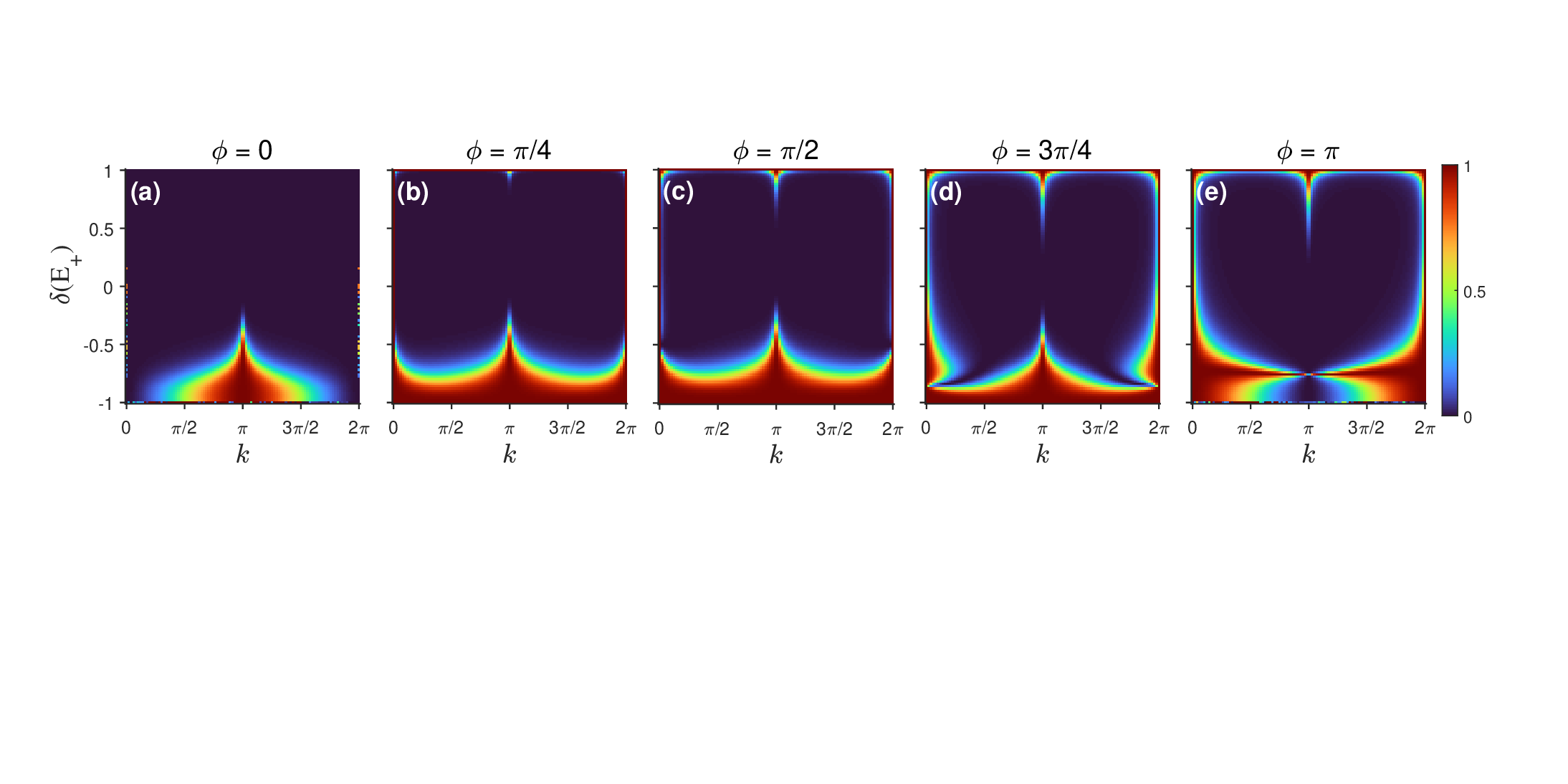}
	\caption{Reflection landscape for stronger gain, $t=1$ and $\gamma=0.3$.}
	\label{fig:R_vs_k_Phic}
\end{figure*}

\begin{figure*}[tbp]
	\centering
	\includegraphics[bb=28 27 995 432, width=18 cm, clip]{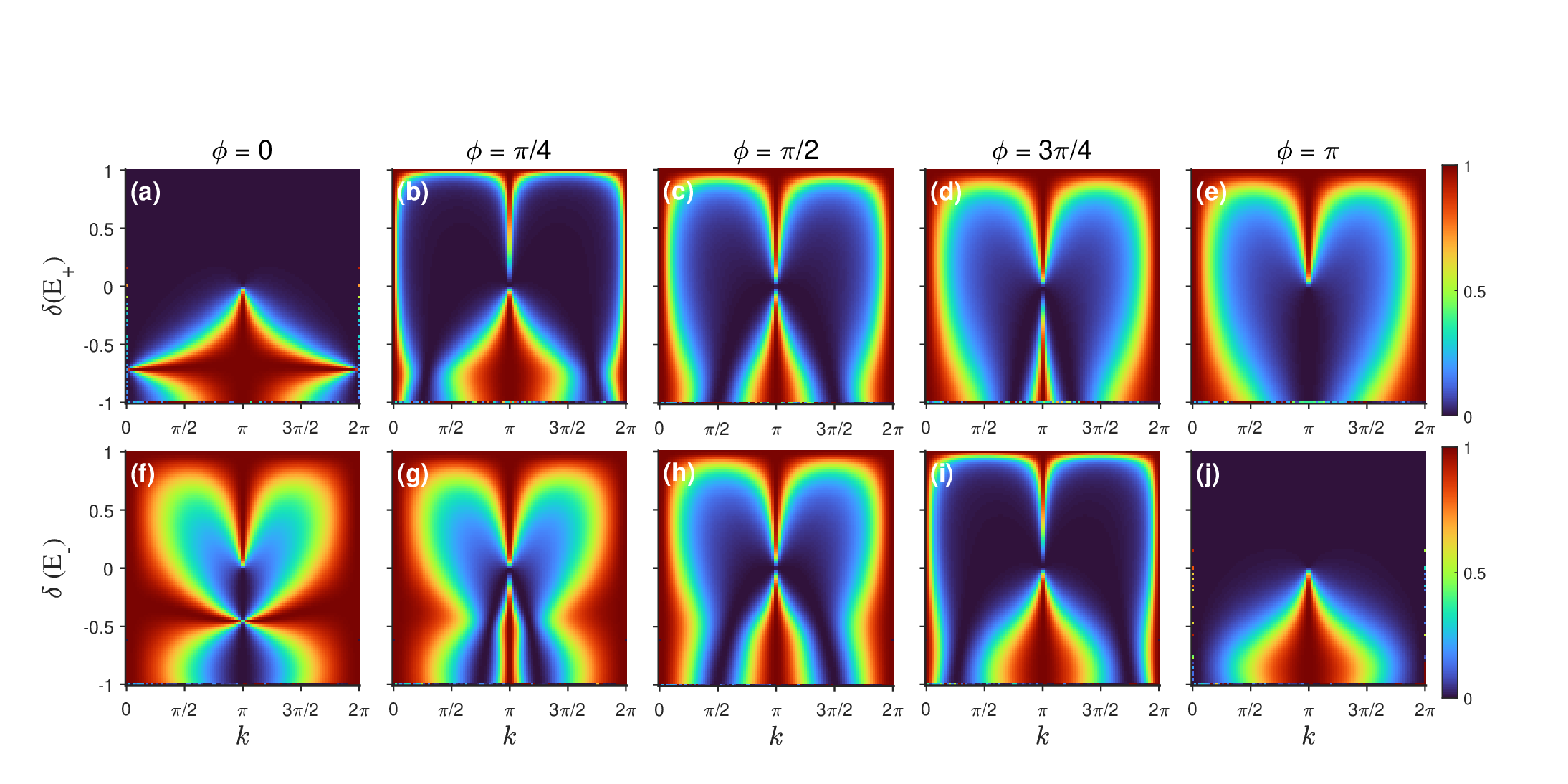}
	\caption{Reflection landscape for weak loss, $t=1$ and $\gamma=-0.1$.}
	\label{fig:R_vs_k_Phid}
\end{figure*}

We now evaluate the analytical expressions for the reflection probability
\begin{equation*}
	R_{L}^{(s)}(E,k,\Phi,\gamma)\equiv |r_{L}^{(s)}(E,k;\Phi,\gamma)|^{2},
\end{equation*}
as a function of crystal momentum $k$ and synthetic phase $\Phi$. Unless stated otherwise, we set $t_{a}=t_{b}=t$ and use the four representative values $\gamma=0$, $0.1$, $0.3$, and $-0.1$. The corresponding reflection maps are shown in Figs.~\ref{fig:R_vs_k_Phia}--\ref{fig:R_vs_k_Phid}.

Several robust features can be read off directly from these maps.

First, the reflection landscapes are miror symmetric about $k=\pi$,
\begin{equation}
	R_{L}^{(s)}(k)=R_{L}^{(s)}(2\pi-k), \label{eq:R_sym}
\end{equation}
which follows from the SSH dispersion and from the way the Bloch phase $\theta(k)$ enters the matching matrix. The symmetry is therefore not accidental: it is inherited from the bulk band geometry and survives the local non-Hermitian perturbation.

Second, gain and loss are not unrelated perturbations. The analytical formulas reveal, and the numerical maps confirm, a useful correspondence between the two once the band index and flux are transformed simultaneously. A representative example is
\begin{equation}
	R_{L}^{(+)}(E_{+},\gamma,\Phi=0)=R_{L}^{(-)}(E_{-},-\gamma,\Phi=\pi), \label{eq:gamma_duality}
\end{equation}
which expresses a loss--gain duality of the reflection problem. Physically, reversing $\gamma$ changes the sign of the qubit self-energy, while shifting $\Phi$ reorganizes the interference between the two coupling paths. Their combined action maps one scattering landscape onto the other.

Third, increasing $|\gamma|$ smooths the sharp resonant structures. This is apparent when Figs.~\ref{fig:R_vs_k_Phia}, \ref{fig:R_vs_k_Phib}, and \ref{fig:R_vs_k_Phic} are compared: the high-contrast reflection features near resonance broaden and lose visibility as the non-Hermitian scale increases. The dominant control parameter for this broadening is the magnitude $|\gamma|$, whereas the sign of $\gamma$ distinguishes amplification from attenuation in the total flux balance.

Finally, the SSH dimerization and the synthetic flux act together as a transport selector. For suitable parameter sets, changing the sign and magnitude of the dimerization drives the system between nearly perfect transmission and nearly perfect reflection, while varying $\Phi$ shifts the interference condition that determines where these extrema occur. In this sense, topology enters the scattering problem not as an abstract band label but as a concrete control knob for the observable reflection landscape.

These stationary maps already capture the main physics, but it is important to verify that they are not an artifact of the frequency-domain treatment. We therefore turn next to real-time wave-packet dynamics and compare the long-time reflected and transmitted weights with the analytical predictions obtained from the matching equations.

\subsection{Wave-packet dynamics and verification of stationary scattering}

To verify that the stationary amplitudes derived from Eq.~\eqref{eq:matching_linear_system} indeed describe the physical scattering process, we simulate the real-time evolution of a single-excitation wave packet in the full SSH--qubit system. This comparison is important because the local qubit term is non-Hermitian: one must check explicitly that the long-time reflected and transmitted weights extracted from dynamics coincide with the frequency-domain result.

The initial state is chosen as a Gaussian packet built from the flux-normalized Bloch modes introduced in Sec.~\ref{sec:group_velocity},
\begin{equation}
	\begin{split}
		|\Psi(0)\rangle &= \sum_{s=\pm}\sum_{k} A_s(k) |\Phi_{s,k}\rangle, \\
		A_s(k) &= \mathcal{N} \exp\left[ -\frac{(k-k_0)^2}{2\sigma_k^2} \right],
	\end{split}
	\label{eq:wavepacket_initial}
\end{equation}
where $k_{0}$ is the central momentum, $\sigma_k$ is the spectral width, and $\mathcal{N}$ fixes normalization. The state then evolves under
\begin{equation}
	H_{\mathrm{tot}}=H_{\mathrm{SSH}}+H_{q}+H_{\mathrm{int}},
\end{equation}
according to
\begin{equation}
	i\frac{d}{dt}|\Psi(t)\rangle = H_{\mathrm{tot}}|\Psi(t)\rangle .
	\label{eq:time_evolution}
\end{equation}

The local population is monitored through
\begin{equation}
	\rho_n(t)=|\langle n|\Psi(t)\rangle|^{2},
\end{equation}
and once the reflected and transmitted packets are spatially separated we define
\begin{equation}
	R(t)=\sum_{n<n_{0}}\rho_n(t),\qquad 
	T(t)=\sum_{n>n_{0}}\rho_n(t),
	\label{eq:RT_wavepacket}
\end{equation}
with $n_{0}$ the cell coupled to the qubit. In the Hermitian limit, $R(\infty)+T(\infty)=1$ is recovered exactly. For finite gain or loss the total outgoing weight is no longer conserved, which is precisely the behavior expected from the non-Hermitian self-energy entering the stationary scattering amplitudes.

Across the parameter range examined here, the asymptotic values extracted from Eq.~\eqref{eq:RT_wavepacket} agree very well with the analytical coefficients $R_{L}^{(s)}=|r_{L}^{(s)}|^{2}$ and $T_{L}^{(s)}=|t_{L}^{(s)}|^{2}$. The dynamics therefore confirm the main stationary picture: the SSH band structure fixes the propagation velocity, while the qubit self-energy and the two-path interference encoded by $\Phi$ determine how much of the incident packet is reflected, transmitted, attenuated, or amplified.

\begin{figure}[tbp]
	\centering
	\includegraphics[bb=16 7 577 431, width=18 cm, clip,width=1\linewidth]{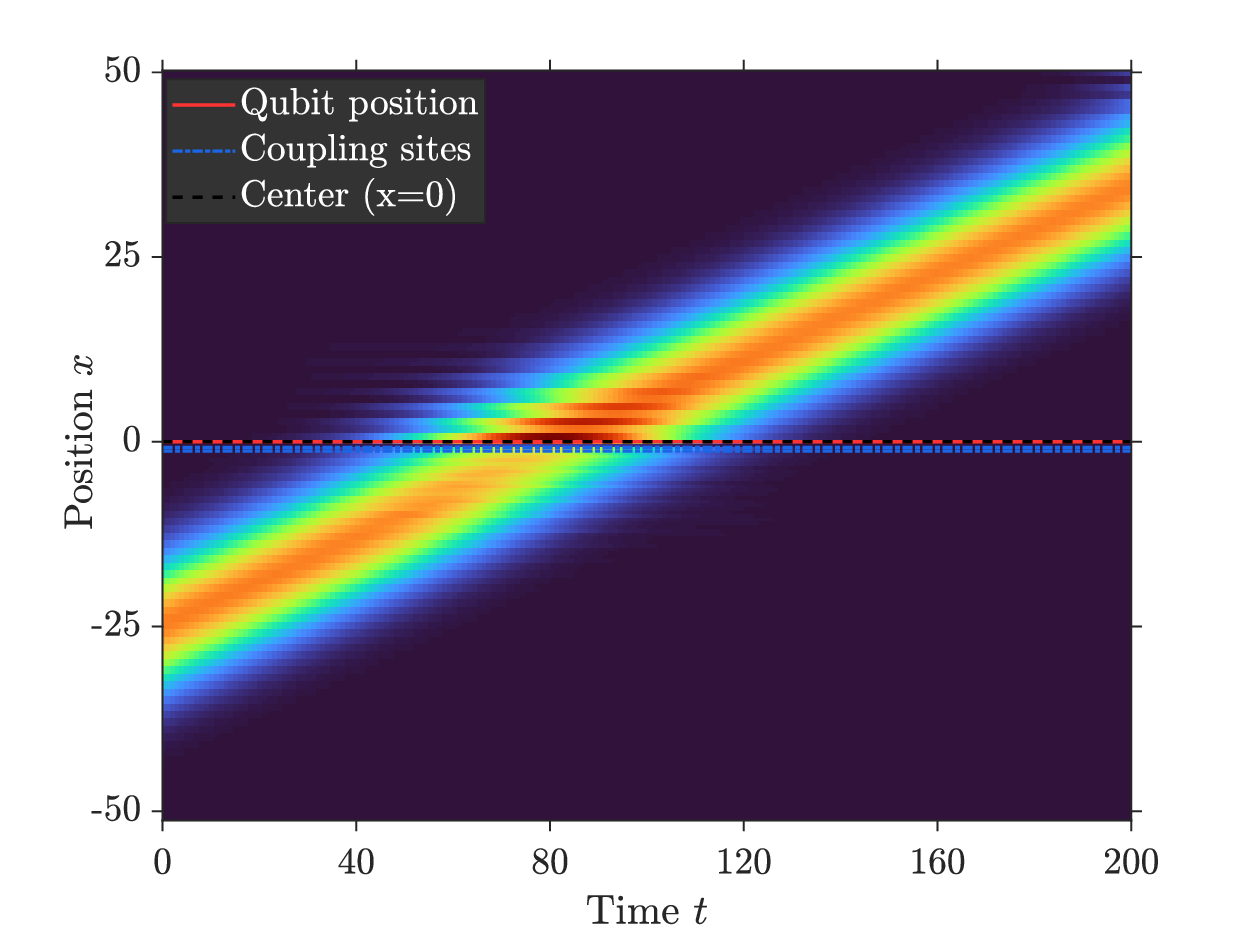}
	\caption{Wave-packet evolution in a transmission-dominated regime for $\delta=0.7$.}
	\label{fig:bobaoT}
\end{figure}

\begin{figure}[tbp]
	\centering
	\includegraphics[bb=16 7 577 431, width=18 cm, clip,width=1\linewidth]{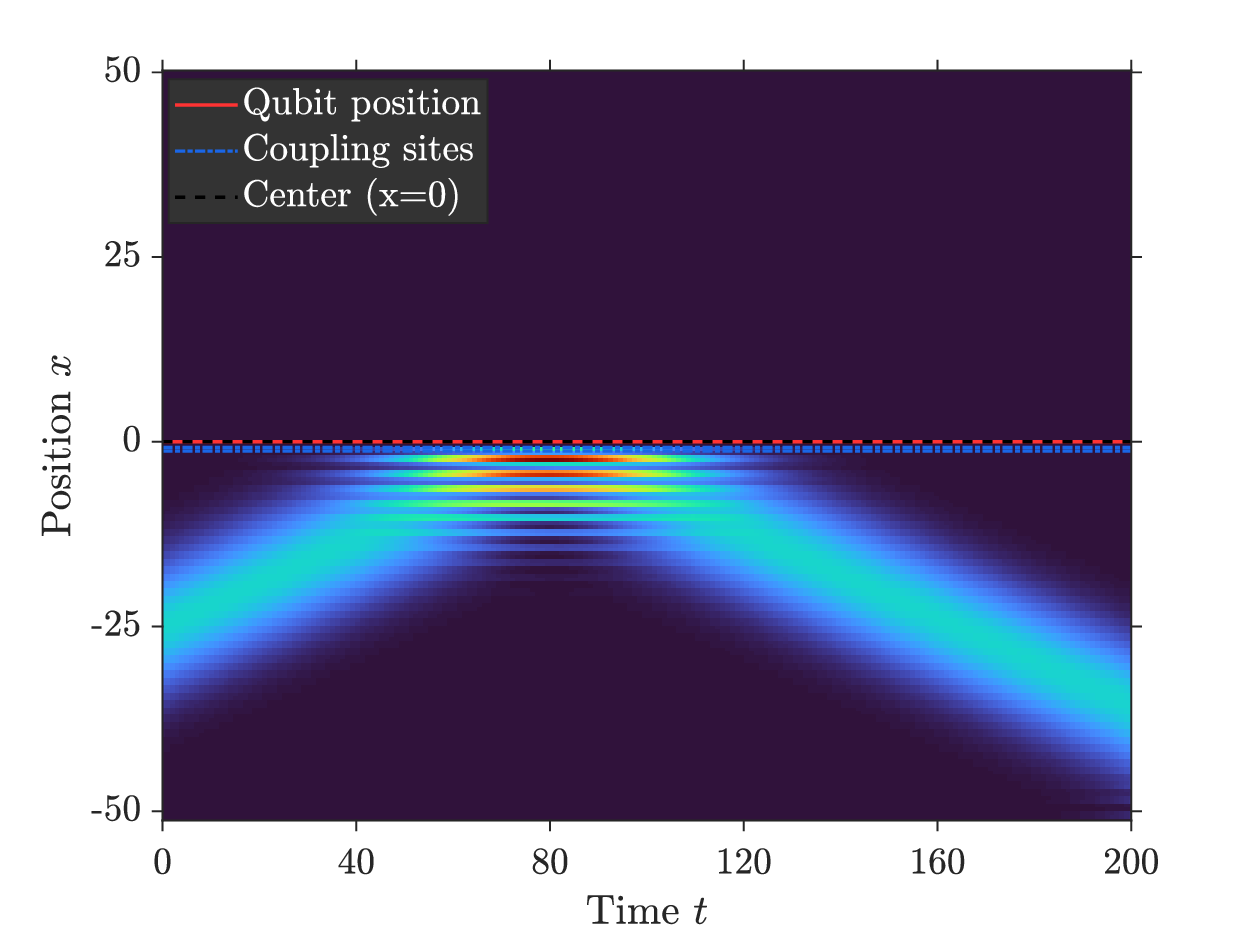}
	\caption{Wave-packet evolution in a reflection-dominated regime for $\delta=-0.7$.}
	\label{fig:bobaoR}
\end{figure}

A particularly clear illustration is obtained at $t_{a}=t_{b}=t=1$, $k_{0}=\pi/2$, $\Phi=0.39$, and $\gamma=1.41$, where tuning the dimerization sharply changes the transport outcome. For $\delta=0.7$, the stationary solution is transmission dominated and the time-evolved packet crosses the scattering region with negligible reflected weight, as shown in Fig.~\ref{fig:bobaoT}. For $\delta=-0.7$, the same protocol yields an almost fully reflected packet, shown in Fig.~\ref{fig:bobaoR}. The key message is that changing the SSH dimerization reorganizes the local interference condition in a way that can nearly switch the device between transparent and reflective operation.

The dynamical results thus provide a direct real-space confirmation of the stationary theory. They also make the physical mechanism more transparent: the qubit does not simply act as a local absorber or emitter, but as an interference center whose effect is filtered by the SSH Bloch structure and by the coupling phase.

\subsection{Single-photon scattering from the right}

The calculation for right incidence is completely analogous. For a photon incoming from the right, the asymptotic SSH amplitudes are
\begin{subequations}
	\begin{align}
		u_{A,k}(n)&=
		\begin{cases}
			t_{R}^{(s)}e^{-ikn}, & n\ll 0, \\[4pt]
			e^{-ikn}+r_{R}^{(s)}e^{ikn}, & n\gg 0,
		\end{cases} \\
		u_{B,k}(n)&=
		\begin{cases}
			s\,t_{R}^{(s)}e^{i\theta(k)}e^{-ikn}, & n\ll 0, \\[4pt]
			s e^{i\theta(k)}e^{-ikn}+s r_{R}^{(s)}e^{-i\theta(k)}e^{ikn}, & n\gg 0,
		\end{cases}
	\end{align}
\end{subequations}
where $r_{R}^{(s)}$ and $t_{R}^{(s)}$ are the right-reflection and right-transmission amplitudes.

Substituting these forms into Eqs.~\eqref{eq:site_e}--\eqref{eq:site_B} again yields a $2\times 2$ linear problem,
\begin{equation}
	M(E_{s})
	\begin{pmatrix}
		r_{R}^{(s)} \\[4pt]
		t_{R}^{(s)}
	\end{pmatrix}
	=\mathbf{b}(E_{s}),  \label{eq:right_linear_system}
\end{equation}
with
\begin{equation*}
	M(E_{s})=
	\begin{pmatrix}
		a_{1} & b_{1} \\[6pt]
		c_{1} & d_{1}
	\end{pmatrix},
	\qquad 
	\mathbf{b}(E_{s})=
	\begin{pmatrix}
		n_{1} \\[4pt]
		n_{2}
	\end{pmatrix},
\end{equation*}
and
\begin{eqnarray*}
	a_{1}=b,\qquad b_{1}=a,\qquad c_{1}=d,\qquad d_{1}=c,\\[4pt]
	n_{1}=\frac{m_{2}}{-s e^{i\theta(k)}},\qquad 
	n_{2}=\frac{m_{1}}{-s e^{i\theta(k)}}.
\end{eqnarray*}
Therefore,
\begin{equation}
	\begin{split}
		r_{R}^{(s)} &= \frac{n_{1}d_{1}-n_{2}b_{1}}{\Delta}, \\
		t_{R}^{(s)} &= \frac{a_{1}n_{2}-c_{1}n_{1}}{\Delta},
	\end{split}
	\qquad 
	\Delta=a_{1}d_{1}-b_{1}c_{1}. \label{eq:rt_right}
\end{equation}
The corresponding probabilities are $R_{R}^{(s)}=|r_{R}^{(s)}|^{2}$ and $T_{R}^{(s)}=|t_{R}^{(s)}|^{2}$.

\begin{figure}[tbp]
	\centering
	\includegraphics[bb=39 12 558 428, width=13 cm,clip,width=1\linewidth]{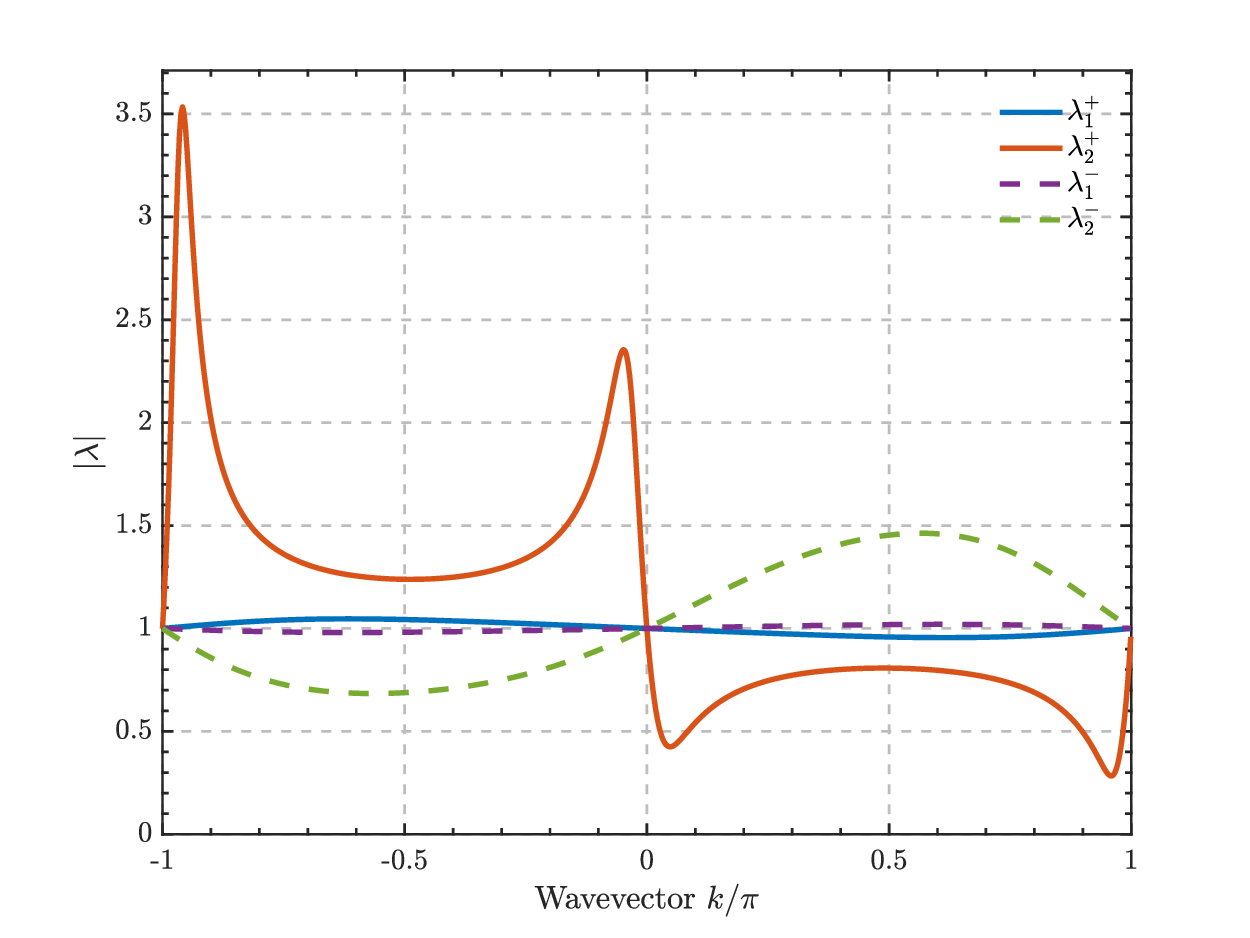}
	\caption{Representative eigenvalue spectrum of the scattering matrix $S(E)$ for the AB coupling geometry.}
	\label{fig:S_K}
\end{figure}

For the parameter sets considered in this work, the reflection spectra for left and right incidence are numerically identical, so the dominant effect of the non-Hermitian qubit is not a strong directional asymmetry in the reflection probability itself but a reorganization of the eigenchannels of the full scattering matrix. Collecting the four amplitudes gives
\begin{equation}
	S(E)=
	\begin{pmatrix}
		r_{L}^{(s)} & t_{R}^{(s)} \\[4pt]
		t_{L}^{(s)} & r_{R}^{(s)}
	\end{pmatrix}.
\end{equation}
Its eigenvalues diagnose attenuation and amplification channel by channel. A representative example is shown in Fig.~\ref{fig:S_K}, where the modulus of the eigenvalues separates dissipative and amplifying sectors in momentum space.

\begin{figure}[t]
	\centering
	\includegraphics[bb=32 101 917 466, width=13 cm,clip,width=1\linewidth]{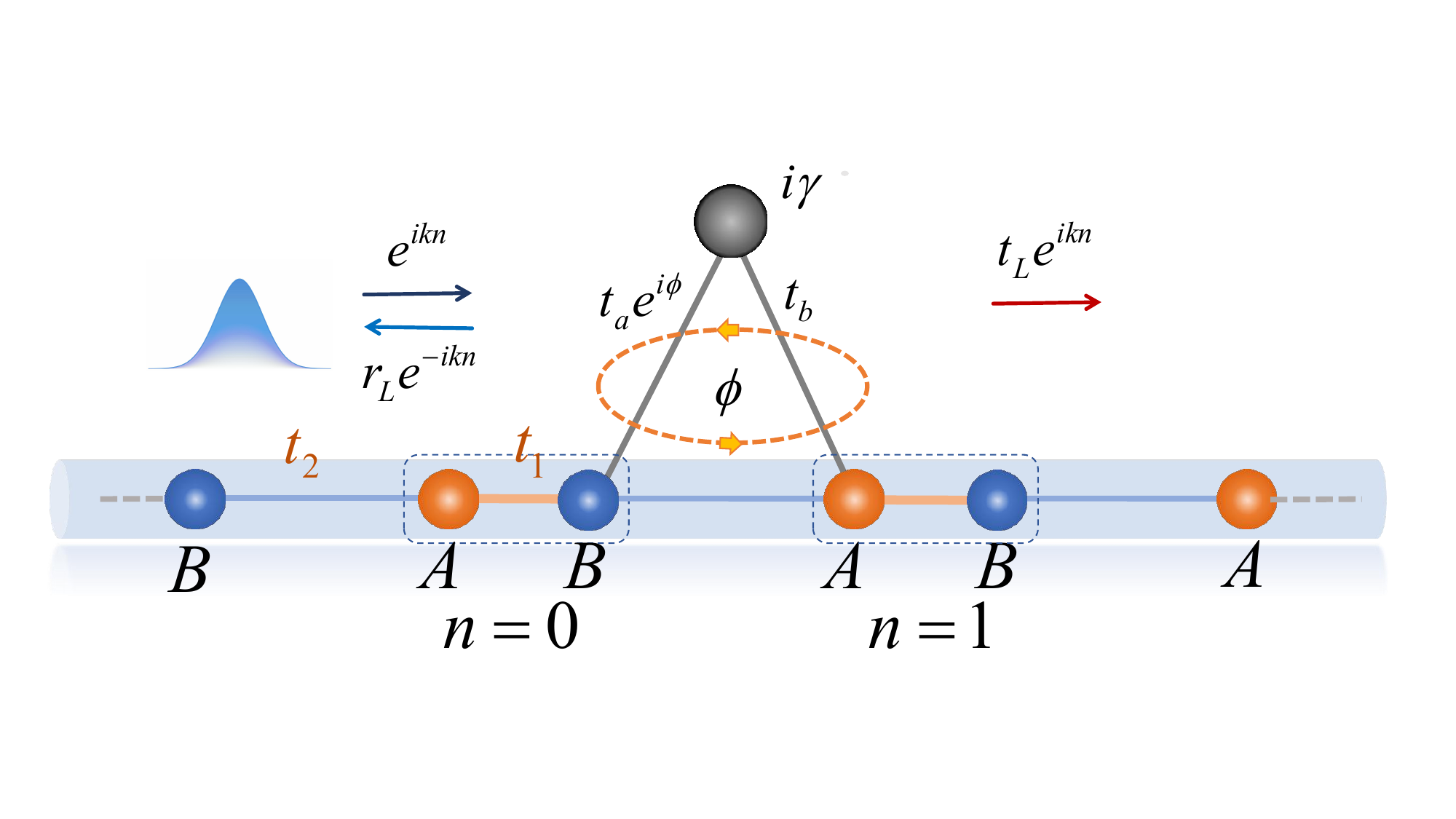}
	\caption{Schematic illustration of a dissipative (or amplifying) superconducting qubit embedded in the center of the SSH lattice. Superconducting qubit simultaneously coupled to the $B$ site of cell $n=0$ and the $A$ site of cell $n=1$.}
	\label{fig:BA}
\end{figure}


\textit{BA coupling geometry.} The same formalism can be extended to an intercell coupling configuration in which the qubit couples to the $B$ site of cell $n=0$ and the $A$ site of cell $n=1$. In that case the interaction Hamiltonian becomes
\begin{equation}
	H_{\mathrm{int}}^{BA}=(t_{a}a_{1}^{\dagger}+t_{b}e^{i\Phi}b_{0}^{\dagger})\sigma^{-}+\mathrm{H.c.},  \label{eq:Hint_BA}
\end{equation}
and the total Hamiltonian is
\begin{align}
	H_{\mathrm{tot}}^{BA}&=H_{\mathrm{SSH}}+H_{q}+H_{\mathrm{int}}^{BA} \notag \\
	&=\sum_{n}[t_{1}a_{n}^{\dagger}b_{n}+t_{2}a_{n+1}^{\dagger}b_{n}+\mathrm{H.c.}] \notag \\
	&\quad+i\gamma |e\rangle \langle e|+[t_{a}e^{i\Phi}b_{0}^{\dagger}+t_{b}a_{1}^{\dagger}\sigma^{-}+\mathrm{H.c.}].  \label{eq:Htot_BA}
\end{align}

Relative to the AB geometry, the BA arrangement inserts an additional propagation segment into the interference loop. As a result, the overall scattering phenomenology remains the same in broad terms, but the locations of resonant extrema in $(k,\Phi)$ space shift and the phase sensitivity becomes more pronounced. In particular, the sign of the SSH dimerization again determines whether the device is driven toward transmission-dominated or reflection-dominated behavior, but the detailed switching boundaries are displaced because the qubit now mediates an intercell rather than an intracell process. The Hermitian reference case for this geometry is shown in Fig.~\ref{fig:BA}.

The BA configuration therefore provides an additional control parameter without changing the basic physical mechanism: topology fixes the Bloch structure, the qubit contributes a local complex self-energy, and the coupling phase determines how the two scattering paths interfere.

\section{Conclusion and Outlook}\label{conclusion}

We have developed a real-space scattering theory for a superconducting qubit coupled locally to an SSH photonic lattice in the presence of tunable loss or gain. By solving the matching problem in the single-excitation sector, we obtained explicit formulas for the reflection and transmission amplitudes for both left and right incidence and organized the response in terms of the eigenvalues of the full scattering matrix. This formulation places coherent perfect absorption, amplification, and ordinary elastic scattering on the same footing.

The central physical result is that topology, local non-Hermiticity, and flux-controlled interference do not act independently. The SSH dimerization fixes the underlying Bloch geometry, the qubit contributes an energy-dependent complex self-energy, and the synthetic phase $\Phi$ selects how the two coupling paths interfere. Together these ingredients can drive the device between transmission-dominated and reflection-dominated regimes and generate highly structured reflection landscapes in momentum and phase space.

The stationary predictions are supported by direct wave-packet simulations. The time-domain dynamics reproduce the long-time reflection and transmission probabilities obtained from the analytic scattering amplitudes and provide a transparent real-space picture of the process: the packet propagates with the SSH group velocity, while the local qubit acts as an interference center whose effect is controlled by the complex self-energy and by the relative phase between the two sublattice channels. We also identified a useful loss--gain correspondence in the reflection maps and found that resonance broadening is governed mainly by the scale $|\gamma|$.

Finally, we showed that the same framework extends naturally from the intracell AB coupling geometry to the intercell BA geometry. The latter does not change the qualitative mechanism, but it shifts the interference condition and offers an additional handle for engineering the scattering response.

These results establish a compact superconducting-circuit platform for non-Hermitian topological scattering. Natural extensions include multiphoton scattering, explicitly driven implementations, non-Markovian qubit environments, and the design of topological absorber or amplifier devices based on scattering-matrix eigenchannels.

\begin{acknowledgments}
	We acknowledge the support of the National Natural Science Foundation of China (Grants No.	12275193, 11975166) and Science \& Technology Development Fund of Tianjin Education Commission for Higher Education(No. 2024KJ059).
\end{acknowledgments}

	\bibliography{QSSH_bibitem}

@article{jalali2023topological,
	title={Topological photonics: Fundamental concepts, recent developments, and future directions},
	author={Jalali Mehrabad, Mahmoud and Mittal, Sunil and Hafezi, Mohammad},
	journal={Phys. Rev. A},
	volume={108},
	number={4},
	pages={040101},
	year={2023},
	publisher={APS},
	doi = {10.1103/PhysRevA.108.040101}
}

@article{zhou2008controllable,
	title={Controllable scattering of a single photon inside a one-dimensional resonator waveguide},
	author={Zhou, Lan and Gong, ZR and Liu, Yu-xi and Sun, CP and Nori, Franco},
	journal={Phys. Rev. Lett.},
	volume={101},
	number={10},
	pages={100501},
	year={2008},
	publisher={APS},
	doi = {10.1103/PhysRevLett.101.100501}
}

@article{yan2021quantum,
	title={Quantum topological photonics},
	author={Yan, Qiuchen and Hu, Xiaoyong and Fu, Yulan and Lu, Cuicui and Fan, Chongxiao and Liu, Qihang and Feng, Xilin and Sun, Quan and Gong, Qihuang},
	journal={Adv. Opt. Mater.},
	volume={9},
	number={15},
	pages={2001739},
	year={2021},
	publisher={Wiley Online Library},
	doi = {10.1002/adom.202001739}
}

@article{RSinglephotonscattering2025,
	title = {Single-photon scattering under control of artificial gauge field},
	author = {Wang, Runting and Wang, Xudong and Mei, Feng and Xiao, Liantuan and Jia, Suotang},
	year = 2025,
	month = feb,
	journal = {Acta Phys. Sin.},
	volume = {74},
	number = {8},
	pages = {084205},
	issn = {0031-9007, 1079-7114},
	doi = {10.7498/aps.74.20250021},
	langid = {english}
}

@article{belloUnconventionalQuantumOptics2019,
  title = {Unconventional Quantum Optics in Topological Waveguide {{QED}}},
  author = {Bello, M. and Platero, G. and Cirac, J. I. and {Gonz{\'a}lez-Tudela}, A.},
  year = 2019,
  month = jul,
  journal = {Sci. Adv.},
  volume = {5},
  number = {7},
  pages = {eaaw0297},
  issn = {2375-2548},
  doi = {10.1126/sciadv.aaw0297},
  langid = {english}
}

@article{benderRealSpectraNonHermitian1998,
  title = {Real {{Spectra}} in {{Non-Hermitian Hamiltonians Having PT-Symmetry}}},
  author = {Bender, Carl M. and Boettcher, Stefan},
  year = 1998,
  month = jun,
  journal = {Phys. Rev. Lett.},
  volume = {80},
  number = {24},
  pages = {5243--5246},
  issn = {0031-9007, 1079-7114},
  doi = {10.1103/PhysRevLett.80.5243},
  copyright = {http://link.aps.org/licenses/aps-default-license},
  langid = {english}
}

@article{blaisCircuitQuantumElectrodynamics2021,
  title = {Circuit Quantum Electrodynamics},
  author = {Blais, Alexandre and Grimsmo, Arne L. and Girvin, S. M. and Wallraff, Andreas},
  year = 2021,
  month = may,
  journal = {Rev. Mod. Phys.},
  volume = {93},
  number = {2},
  pages = {025005},
  issn = {0034-6861, 1539-0756},
  doi = {10.1103/RevModPhys.93.025005},
  langid = {english}
}

@article{chongPTsymmetryBreakingLaserabsorber2010,
  title = {{{PT-symmetry}} Breaking and Laser-Absorber Modes in Optical Scattering Systems},
  author={Chong, YD and Ge, Li and Stone, A Douglas},
  journal={Phys. Rev. Lett.},
  volume={106},
  number={9},
  pages={093902},
  year={2011},
  publisher={APS},
  doi = {10.1103/PhysRevLett.106.093902}
}

@article{dongWaveguideQEDDissipative2025,
  title = {Waveguide {{QED}} with Dissipative Light-Matter Couplings},
  author = {Dong, Xing-Liang and Li, Peng-Bo and Gong, Zongping and Nori, Franco},
  year = 2025,
  month = feb,
  journal = {Phys. Rev. Res.},
  volume = {7},
  number = {1},
  pages = {L012036},
  issn = {2643-1564},
  doi = {10.1103/PhysRevResearch.7.L012036},
  langid = {english}
}

@article{fengNonreciprocalLightPropagation2011,
  title = {Nonreciprocal {{Light Propagation}} in a {{Silicon Photonic Circuit}}},
  author = {Feng, Liang and Ayache, Maurice and Huang, Jingqing and Xu, Ye-Long and Lu, Ming-Hui and Chen, Yan-Feng and Fainman, Yeshaiahu and Scherer, Axel},
  year = 2011,
  month = aug,
  journal = {Science},
  volume = {333},
  number = {6043},
  pages = {729--733},
  issn = {0036-8075, 1095-9203},
  doi = {10.1126/science.1206038},
  langid = {english}
}

@article{gaoQuantumTopologicalPhotonics2024,
  title = {Quantum Topological Photonics with Special Focus on Waveguide Systems},
  author = {Gao, Jun and Xu, Ze-Sheng and Yang, Zhaoju and Zwiller, Val and Elshaari, Ali W.},
  year = 2024,
  month = aug,
  journal = {npj Nanophoton},
  volume = {1},
  number = {1},
  pages = {34},
  issn = {2948-216X},
  doi = {10.1038/s44310-024-00034-5},
  langid = {english}
}

@article{geConservationRelationsAnisotropic2012,
  title = {Conservation Relations and Anisotropic Transmission Resonances in One-Dimensional {{PT}}-Symmetric Photonic Heterostructures},
  author = {Ge, Li and Chong, Y. D. and Stone, A. D.},
  year = 2012,
  month = feb,
  journal = {Phys. Rev. A},
  volume = {85},
  number = {2},
  pages = {023802},
  issn = {1050-2947, 1094-1622},
  doi = {10.1103/PhysRevA.85.023802},
  copyright = {https://doi.org/10.1103/PhysRevA.85.023802},
  langid = {english}
}

@article{geContrastingEigenvalueSingularvalue2017,
  title = {Contrasting Eigenvalue and Singular-Value Spectra for Lasing and Antilasing in a {{PT}}-Symmetric Periodic Structure},
  author = {Ge, Li and Feng, Liang},
  year = 2017,
  month = jan,
  journal = {Phys. Rev. A},
  volume = {95},
  number = {1},
  pages = {013813},
  issn = {2469-9926, 2469-9934},
  doi = {10.1103/PhysRevA.95.013813},
  copyright = {http://doi.org/10.1103/PhysRevA.95.013813},
  langid = {english}
}

@article{hafeziImagingTopologicalEdge2013,
  title = {Imaging Topological Edge States in Silicon Photonics},
  author = {Hafezi, M. and Mittal, S. and Fan, J. and Migdall, A. and Taylor, J. M.},
  year = 2013,
  month = dec,
  journal = {Nat. Photon.},
  volume = {7},
  number = {12},
  pages = {1001--1005},
  issn = {1749-4885, 1749-4893},
  doi = {10.1038/nphoton.2013.274},
  copyright = {https://doi.org/10.1038/nphoton.2013.274},
  langid = {english}
}

@article{jinUnitaryScatteringProtected2022,
  title = {Unitary {{Scattering Protected}} by {{Pseudo-Hermiticity}}},
  author = {Jin, L.},
  year = 2022,
  month = feb,
  journal = {Chin. Phys. Lett.},
  volume = {39},
  number = {3},
  pages = {037302},
  issn = {0256-307X, 1741-3540},
  doi = {10.1088/0256-307X/39/3/037302}
}

@article{khurginNonreciprocalPropagationNonreciprocal2020,
  title = {Non-Reciprocal Propagation versus Non-Reciprocal Control},
  author = {Khurgin, Jacob B.},
  year = 2020,
  month = dec,
  journal = {Nat. Photon.},
  volume = {14},
  number = {12},
  pages = {711--711},
  issn = {1749-4885, 1749-4893},
  doi = {10.1038/s41566-020-00723-5},
  langid = {english}
}

@article{kimQuantumElectrodynamicsTopological2021a,
  title = {Quantum {{Electrodynamics}} in a {{Topological Waveguide}}},
  author = {Kim, Eunjong and Zhang, Xueyue and Ferreira, Vinicius S. and Banker, Jash and Iverson, Joseph K. and Sipahigil, Alp and Bello, Miguel and {Gonz{\'a}lez-Tudela}, Alejandro and Mirhosseini, Mohammad and Painter, Oskar},
  year = 2021,
  month = jan,
  journal = {Phys. Rev. X},
  volume = {11},
  number = {1},
  pages = {011015},
  issn = {2160-3308},
  doi = {10.1103/PhysRevX.11.011015},
  langid = {english}
}

@article{kjaergaardSuperconductingQubitsCurrent2020,
  title = {Superconducting {{Qubits}}: {{Current State}} of {{Play}}},
  shorttitle = {Superconducting {{Qubits}}},
  author = {Kjaergaard, Morten and Schwartz, Mollie E. and Braum{\"u}ller, Jochen and Krantz, Philip and Wang, Joel I.-J. and Gustavsson, Simon and Oliver, William D.},
  year = 2020,
  month = mar,
  journal = {Annu. Rev. Condens. Matter Phys.},
  volume = {11},
  number = {1},
  pages = {369--395},
  issn = {1947-5454, 1947-5462},
  doi = {10.1146/annurev-conmatphys-031119-050605},
  langid = {english}
}

@article{linMeasuringNonHermitianTopological2025,
  title = {Measuring Non-{{Hermitian}} Topological Invariants Directly from Quench Dynamics},
  author = {Lin, Xiao-Dong and Zhang, Long},
  year = 2025,
  month = mar,
  journal = {Phys. Rev. Res.},
  volume = {7},
  number = {1},
  pages = {L012060},
  issn = {2643-1564},
  doi = {10.1103/PhysRevResearch.7.L012060},
  langid = {english}
}

@article{longhiPTSymmetricLaser2010,
  title = {{{PT}} -Symmetric Laser Absorber},
  author = {Longhi, Stefano},
  year = 2010,
  month = sep,
  journal = {Phys. Rev. A},
  volume = {82},
  number = {3},
  pages = {031801},
  issn = {1050-2947, 1094-1622},
  doi = {10.1103/PhysRevA.82.031801},
  langid = {english}
}

@article{luDynamicalTopologyChiral2025,
  title = {Dynamical Topology of Chiral and Nonreciprocal State Transfers in a Non-Hermitian Quantum System},
  author = {Lu, Pengfei and Liu, Yang and Lao, Qifeng and Liu, Teng and Rao, Xinxin and Bian, Ji and Wu, Hao and Zhu, Feng and Luo, Le},
  year = 2025,
  month = mar,
  journal = {Commun. Phys.},
  volume = {8},
  number = {1},
  pages = {91},
  issn = {2399-3650},
  doi = {10.1038/s42005-025-01989-3},
  langid = {english}
}

@article{luTopologicalPhotonics2014a,
  title = {Topological Photonics},
  author = {Lu, Ling and Joannopoulos, John D. and Solja{\v c}i{\'c}, Marin},
  year = 2014,
  month = nov,
  journal = {Nat. Photon.},
  volume = {8},
  number = {11},
  pages = {821--829},
  issn = {1749-4885, 1749-4893},
  doi = {10.1038/nphoton.2014.248},
  langid = {english}
}

@article{meierObservationTopologicalSoliton2016,
  title = {Observation of the Topological Soliton State in the {{Su}}--{{Schrieffer}}--{{Heeger}} Model},
  author = {Meier, Eric J. and An, Fangzhao Alex and Gadway, Bryce},
  year = 2016,
  month = dec,
  journal = {Nat. Commun.},
  volume = {7},
  number = {1},
  pages = {13986},
  issn = {2041-1723},
  doi = {10.1038/ncomms13986},
  langid = {english}
}

@article{mondalSuSchriefferHeegerModelFundamentals2025,
  title = {Su-{{Schrieffer-Heeger Model}}- {{From Fundamentals}} to {{Responses}}},
  author = {Mondal, Deep and Bandyopadhyay, Arka and Jana, Debnarayan},
  year = 2025,
  month = apr,
  journal = {Int. J. Theor. Phys.},
  volume = {64},
  number = {5},
  pages = {125},
  issn = {1572-9575},
  doi = {10.1007/s10773-025-05981-z},
  langid = {english}
}

@article{mostafazadehInvisibilityPTSymmetry2013,
  title = {Invisibility and {{PT}}-Symmetry},
  author = {Mostafazadeh, Ali},
  year = 2013,
  month = jan,
  journal = {Phys. Rev. A},
  volume = {87},
  number = {1},
  pages = {012103},
  issn = {1050-2947, 1094-1622},
  doi = {10.1103/PhysRevA.87.012103},
  langid = {english}
}

@article{mostafazadehSpectralSingularitiesComplex2009,
  title = {Spectral {{Singularities}} of {{Complex Scattering Potentials}} and {{Infinite Reflection}} and {{Transmission Coefficients}} at {{Real Energies}}},
  author = {Mostafazadeh, Ali},
  year = 2009,
  month = jun,
  journal = {Phys. Rev. Lett.},
  volume = {102},
  number = {22},
  pages = {220402},
  issn = {0031-9007, 1079-7114},
  doi = {10.1103/PhysRevLett.102.220402},
  langid = {english}
}

@article{ozawaTopologicalPhotonics2019,
  title = {Topological Photonics},
  author = {Ozawa, Tomoki and Price, Hannah M. and Amo, Alberto and Goldman, Nathan and Hafezi, Mohammad and Lu, Ling and Rechtsman, Mikael C. and Schuster, David and Simon, Jonathan and Zilberberg, Oded and Carusotto, Iacopo},
  year = 2019,
  month = mar,
  journal = {Rev. Mod. Phys.},
  volume = {91},
  number = {1},
  pages = {015006},
  issn = {0034-6861, 1539-0756},
  doi = {10.1103/RevModPhys.91.015006},
  langid = {english}
}

@article{piasotskiDiagrammaticApproachScattering2021,
  title ={Diagrammatic {{Approach}} to {{Scattering}} of {{Multi-Photon States}} in {{Waveguide QED}}},
  author={Piasotski, Kiryl and Pletyukhov, Mikhail},
  journal={Phys. Rev. A},
  volume={104},
  number={2},
  pages={023709},
  year={2021},
  publisher={APS},
  doi = {10.1103/PhysRevA.104.023709}
}

@article{poddubnyVanWaalsMaterials2025,
  title = {Van Der {{Waals}} Materials for Waveguide {{QED}}},
  author = {Poddubny, Alexander},
  year = 2025,
  month = aug,
  journal = {Nat. Photon.},
  volume = {19},
  number = {8},
  pages = {783--784},
  issn = {1749-4885, 1749-4893},
  doi = {10.1038/s41566-025-01726-w},
  langid = {english}
}

@article{ramezaniNonreciprocalLocalizationPhotons2018,
  title = {Nonreciprocal {{Localization}} of {{Photons}}},
  author = {Ramezani, Hamidreza and Jha, Pankaj K. and Wang, Yuan and Zhang, Xiang},
  year = 2018,
  month = jan,
  journal = {Phys. Rev. Lett.},
  volume = {120},
  number = {4},
  pages = {043901},
  issn = {0031-9007, 1079-7114},
  doi = {10.1103/PhysRevLett.120.043901},
  langid = {english}
}

@article{riveroTimereversalinvariantScalingLight2019,
  title = {Time-Reversal-Invariant Scaling of Light Propagation in One-Dimensional Non-{{Hermitian}} Systems},
  author = {Rivero, Jose D. H. and Ge, Li},
  year = 2019,
  month = aug,
  journal = {Phys. Rev. A},
  volume = {100},
  number = {2},
  pages = {023819},
  issn = {2469-9926, 2469-9934},
  doi = {10.1103/PhysRevA.100.023819},
  langid = {english}
}

@article{sauerTheoryIntrinsicPropagation2020,
  title = {Theory of Intrinsic Propagation Losses in Topological Edge States of Planar Photonic Crystals},
  author = {Sauer, Erik and Vasco, Juan Pablo and Hughes, Stephen},
  year = 2020,
  month = oct,
  journal = {Phys. Rev. Res.},
  volume = {2},
  number = {4},
  pages = {043109},
  issn = {2643-1564},
  doi = {10.1103/PhysRevResearch.2.043109},
  langid = {english}
}

@article{wanTimeReversedLasingInterferometric2011,
  title = {Time-{{Reversed Lasing}} and {{Interferometric Control}} of {{Absorption}}},
  author = {Wan, Wenjie and Chong, Yidong and Ge, Li and Noh, Heeso and Stone, A. Douglas and Cao, Hui},
  year = 2011,
  month = feb,
  journal = {Science},
  volume = {331},
  number = {6019},
  pages = {889--892},
  issn = {0036-8075, 1095-9203},
  doi = {10.1126/science.1200735},
  langid = {english}
}

@article{warnerCoherentControlSuperconducting2025,
  title = {Coherent Control of a Superconducting Qubit Using Light},
  author = {Warner, Hana K. and Holzgrafe, Jeffrey and Yankelevich, Beatriz and Barton, David and Poletto, Stefano and Xin, C. J. and Sinclair, Neil and Zhu, Di and Sete, Eyob and Langley, Brandon and Batson, Emma and Colangelo, Marco and {Shams-Ansari}, Amirhassan and Joe, Graham and Berggren, Karl K. and Jiang, Liang and Reagor, Matthew J. and Lon{\v c}ar, Marko},
  year = 2025,
  month = may,
  journal = {Nat. Phys.},
  volume = {21},
  number = {5},
  pages = {831--838},
  issn = {1745-2473, 1745-2481},
  doi = {10.1038/s41567-025-02812-0},
  langid = {english}
}

@article{youssefiTopologicalLatticesRealized2022,
  title = {Topological Lattices Realized in Superconducting Circuit Optomechanics},
  author = {Youssefi, Amir and Kono, Shingo and Bancora, Andrea and Chegnizadeh, Mahdi and Pan, Jiahe and Vovk, Tatiana and Kippenberg, Tobias J.},
  year = 2022,
  month = dec,
  journal = {Nature},
  volume = {612},
  number = {7941},
  pages = {666--672},
  issn = {0028-0836, 1476-4687},
  doi = {10.1038/s41586-022-05367-9},
  langid = {english}
}

@article{zhuSinglephotonScatteringGiantatom2025a,
  title = {Single-Photon Scattering in Giant-Atom Topological-Waveguide-{{QED}} Systems},
  author = {Zhu, Hai and Yin, Xian-Li and Liao, Jie-Qiao},
  year = 2025,
  month = feb,
  journal = {Phys. Rev. A},
  volume = {111},
  number = {2},
  pages = {023711},
  issn = {2469-9926, 2469-9934},
  doi = {10.1103/PhysRevA.111.023711},
  langid = {english}
}

\end{document}